\documentclass[a4paper, amsfonts, amssymb, amsmath, reprint, showkeys, nofootinbib, twoside,superscriptaddress,floatfix]{revtex4-1}
\usepackage[english]{babel}
\usepackage[utf8]{inputenc}
\usepackage[colorinlistoftodos, color=green!40, prependcaption]{todonotes}

\usepackage{amsthm}
\usepackage{mathtools}
\usepackage{physics}
\usepackage{xcolor}
\usepackage{graphicx}
\usepackage[left=23mm,right=13mm,top=35mm,columnsep=15pt]{geometry} 
\usepackage{adjustbox}
\usepackage{placeins}
\usepackage[T1]{fontenc}
\usepackage{lipsum}
\usepackage{csquotes}
\usepackage{amsmath}
\usepackage{amssymb}
\usepackage{amsfonts}
\usepackage{tikz}
\def\mT{\mathcal{T}}
\def\mP{\mathcal{P}}
\def\mS{\mathcal{S}}

\def\Fmovebasisa#1#2#3#4#5{
\tikzset{every picture/.style={line width=0.75pt}} 
\begin{tikzpicture}[x=0.75pt,y=0.75pt,yscale=-0.5,xscale=0.5,baseline=(current bounding box.center)]
\draw    (320,220) -- (320,202) ;
\draw [shift={(320,200)}, rotate = 450] [color={rgb, 255:red, 0; green, 0; blue, 0 }  ][line width=0.75]    (10.93,-3.29) .. controls (6.95,-1.4) and (3.31,-0.3) .. (0,0) .. controls (3.31,0.3) and (6.95,1.4) .. (10.93,3.29)   ;
\draw    (320,180) -- (320,200) ;
\draw    (280,140) -- (298.59,121.41) ;
\draw [shift={(300,120)}, rotate = 495] [color={rgb, 255:red, 0; green, 0; blue, 0 }  ][line width=0.75]    (10.93,-3.29) .. controls (6.95,-1.4) and (3.31,-0.3) .. (0,0) .. controls (3.31,0.3) and (6.95,1.4) .. (10.93,3.29)   ;
\draw    (300,120) -- (320,100) ;
\draw    (280,140) -- (300,160) ;
\draw    (320,180) -- (301.41,161.41) ;
\draw [shift={(300,160)}, rotate = 405] [color={rgb, 255:red, 0; green, 0; blue, 0 }  ][line width=0.75]    (10.93,-3.29) .. controls (6.95,-1.4) and (3.31,-0.3) .. (0,0) .. controls (3.31,0.3) and (6.95,1.4) .. (10.93,3.29)   ;
\draw    (240,100) -- (260,120) ;
\draw    (280,140) -- (261.41,121.41) ;
\draw [shift={(260,120)}, rotate = 405] [color={rgb, 255:red, 0; green, 0; blue, 0 }  ][line width=0.75]    (10.93,-3.29) .. controls (6.95,-1.4) and (3.31,-0.3) .. (0,0) .. controls (3.31,0.3) and (6.95,1.4) .. (10.93,3.29)   ;
\draw    (320,180) -- (358.59,141.41) ;
\draw [shift={(360,140)}, rotate = 495] [color={rgb, 255:red, 0; green, 0; blue, 0 }  ][line width=0.75]    (10.93,-3.29) .. controls (6.95,-1.4) and (3.31,-0.3) .. (0,0) .. controls (3.31,0.3) and (6.95,1.4) .. (10.93,3.29)   ;
\draw    (400,100) -- (360,140) ;
\draw (320,97) node [anchor=south] [inner sep=0.75pt]   [align=left] {$#2$};
\draw (322,200) node [anchor=west] [inner sep=0.75pt]   [align=left] {$#4$};
\draw (400,97) node [anchor=south] [inner sep=0.75pt]   [align=left] {$#3$};
\draw (240,97) node [anchor=south] [inner sep=0.75pt]   [align=left] {$#1$};
\draw (298,163) node [anchor=north east] [inner sep=0.75pt]   [align=left] {$#5$};
\end{tikzpicture}
}

\def\Fmovebasisb#1#2#3#4#5{
\tikzset{every picture/.style={line width=0.75pt}} 
\begin{tikzpicture}[x=0.75pt,y=0.75pt,yscale=-0.5,xscale=0.5,baseline=(current bounding box.center)]
\draw    (320,220) -- (320,202) ;
\draw [shift={(320,200)}, rotate = 450] [color={rgb, 255:red, 0; green, 0; blue, 0 }  ][line width=0.75]    (10.93,-3.29) .. controls (6.95,-1.4) and (3.31,-0.3) .. (0,0) .. controls (3.31,0.3) and (6.95,1.4) .. (10.93,3.29)   ;
\draw    (320,180) -- (320,200) ;
\draw    (240,100) -- (280,140) ;
\draw    (320,180) -- (281.41,141.41) ;
\draw [shift={(280,140)}, rotate = 405] [color={rgb, 255:red, 0; green, 0; blue, 0 }  ][line width=0.75]    (10.93,-3.29) .. controls (6.95,-1.4) and (3.31,-0.3) .. (0,0) .. controls (3.31,0.3) and (6.95,1.4) .. (10.93,3.29)   ;
\draw    (400,100) -- (380,120) ;
\draw    (360,140) -- (378.59,121.41) ;
\draw [shift={(380,120)}, rotate = 495] [color={rgb, 255:red, 0; green, 0; blue, 0 }  ][line width=0.75]    (10.93,-3.29) .. controls (6.95,-1.4) and (3.31,-0.3) .. (0,0) .. controls (3.31,0.3) and (6.95,1.4) .. (10.93,3.29)   ;
\draw    (360,140) -- (340,160) ;
\draw    (320,180) -- (338.59,161.41) ;
\draw [shift={(340,160)}, rotate = 495] [color={rgb, 255:red, 0; green, 0; blue, 0 }  ][line width=0.75]    (10.93,-3.29) .. controls (6.95,-1.4) and (3.31,-0.3) .. (0,0) .. controls (3.31,0.3) and (6.95,1.4) .. (10.93,3.29)   ;
\draw    (360,140) -- (341.41,121.41) ;
\draw [shift={(340,120)}, rotate = 405] [color={rgb, 255:red, 0; green, 0; blue, 0 }  ][line width=0.75]    (10.93,-3.29) .. controls (6.95,-1.4) and (3.31,-0.3) .. (0,0) .. controls (3.31,0.3) and (6.95,1.4) .. (10.93,3.29)   ;
\draw    (320,100) -- (340,120) ;
\draw (322,200) node [anchor=west] [inner sep=0.75pt]   [align=left] {$#4$};
\draw (400,97) node [anchor=south] [inner sep=0.75pt]   [align=left] {$#3$};
\draw (240,97) node [anchor=south] [inner sep=0.75pt]   [align=left] {$#1$};
\draw (320,97) node [anchor=south] [inner sep=0.75pt]   [align=left] {$#2$};
\draw (348,162) node [anchor=north west][inner sep=0.75pt]   [align=left] {$#5$};
\end{tikzpicture}
}

\def\Bpselfdual#1#2#3#4#5#6{
\tikzset{every picture/.style={line width=0.75pt}} 
\begin{tikzpicture}[x=0.75pt,y=0.75pt,yscale=-0.5,xscale=0.5,baseline=(current bounding box.center)]
\draw    (360,160) -- (360,120) ;
\draw    (280,120) -- (280,160) ;
\draw [color={rgb, 255:red, 0; green, 0; blue, 0 }  ,draw opacity=1 ]   (280,160) -- (240,200) ;
\draw    (280,160) -- (320,200) ;
\draw    (320,240) -- (320,200) ;
\draw    (360,160) -- (320,200) ;
\draw    (360,160) -- (400,200) ;
\draw    (320,80) -- (280,120) ;
\draw    (240,80) -- (280,120) ;
\draw    (400,80) -- (360,120) ;
\draw    (320,80) -- (360,120) ;
\draw    (320,80) -- (320,40) ;
\draw (318,60) node [anchor=east] [inner sep=0.75pt]   [align=left] {$\displaystyle a$};
\draw (378,97) node [anchor=south east] [inner sep=0.75pt]   [align=left] {$\displaystyle b$};
\draw (382,177) node [anchor=south west] [inner sep=0.75pt]   [align=left] {$\displaystyle c$};
\draw (322,220) node [anchor=west] [inner sep=0.75pt]   [align=left] {$\displaystyle d$};
\draw (262,183) node [anchor=north west][inner sep=0.75pt]   [align=left] {$\displaystyle e$};
\draw (258,103) node [anchor=north east] [inner sep=0.75pt]   [align=left] {$\displaystyle f$};
\draw (362,140) node [anchor=west] [inner sep=0.75pt]   [align=left] {$\displaystyle #2$};
\draw (338,177) node [anchor=south east] [inner sep=0.75pt]   [align=left] {$\displaystyle #3$};
\draw (298,183) node [anchor=north east] [inner sep=0.75pt]   [align=left] {$\displaystyle #4$};
\draw (282,140) node [anchor=west] [inner sep=0.75pt]   [align=left] {$\displaystyle #5$};
\draw (298,97) node [anchor=south east] [inner sep=0.75pt]   [align=left] {$\displaystyle #6$};
\draw (338,103) node [anchor=north east] [inner sep=0.75pt]   [align=left] {$\displaystyle #1$};
\end{tikzpicture}
}

\def\torus#1#2#3{
\tikzset{every picture/.style={line width=0.75pt}} 
\begin{tikzpicture}[x=0.75pt,y=0.75pt,yscale=-0.5,xscale=0.5,baseline=(current bounding box.center)]
\draw    (340,60) -- (400,60) ;
\draw    (400,60) -- (400,240) ;
\draw    (220,60) -- (220,240) ;
\draw    (220,120) -- (280,120) ;
\draw    (280,120) -- (340,180) ;
\draw    (340,180) -- (400,180) ;
\draw    (280,120) -- (300,60) ;
\draw    (320,240) -- (340,180) ;
\draw    (220,60) -- (278,60) ;
\draw [shift={(280,60)}, rotate = 180] [color={rgb, 255:red, 0; green, 0; blue, 0 }  ][line width=0.75]    (10.93,-3.29) .. controls (6.95,-1.4) and (3.31,-0.3) .. (0,0) .. controls (3.31,0.3) and (6.95,1.4) .. (10.93,3.29)   ;
\draw    (280,60) -- (338,60) ;
\draw [shift={(340,60)}, rotate = 180] [color={rgb, 255:red, 0; green, 0; blue, 0 }  ][line width=0.75]    (10.93,-3.29) .. controls (6.95,-1.4) and (3.31,-0.3) .. (0,0) .. controls (3.31,0.3) and (6.95,1.4) .. (10.93,3.29)   ;
\draw    (340,240) -- (400,240) ;
\draw    (280,240) -- (338,240) ;
\draw [shift={(340,240)}, rotate = 180] [color={rgb, 255:red, 0; green, 0; blue, 0 }  ][line width=0.75]    (10.93,-3.29) .. controls (6.95,-1.4) and (3.31,-0.3) .. (0,0) .. controls (3.31,0.3) and (6.95,1.4) .. (10.93,3.29)   ;
\draw    (220,240) -- (278,240) ;
\draw [shift={(280,240)}, rotate = 180] [color={rgb, 255:red, 0; green, 0; blue, 0 }  ][line width=0.75]    (10.93,-3.29) .. controls (6.95,-1.4) and (3.31,-0.3) .. (0,0) .. controls (3.31,0.3) and (6.95,1.4) .. (10.93,3.29)   ;
\draw    (220,160) -- (220,102) ;
\draw [shift={(220,100)}, rotate = 450] [color={rgb, 255:red, 0; green, 0; blue, 0 }  ][line width=0.75]    (10.93,-3.29) .. controls (6.95,-1.4) and (3.31,-0.3) .. (0,0) .. controls (3.31,0.3) and (6.95,1.4) .. (10.93,3.29)   ;
\draw    (220,180) -- (220,142) ;
\draw [shift={(220,140)}, rotate = 450] [color={rgb, 255:red, 0; green, 0; blue, 0 }  ][line width=0.75]    (10.93,-3.29) .. controls (6.95,-1.4) and (3.31,-0.3) .. (0,0) .. controls (3.31,0.3) and (6.95,1.4) .. (10.93,3.29)   ;
\draw    (400,160) -- (400,102) ;
\draw [shift={(400,100)}, rotate = 450] [color={rgb, 255:red, 0; green, 0; blue, 0 }  ][line width=0.75]    (10.93,-3.29) .. controls (6.95,-1.4) and (3.31,-0.3) .. (0,0) .. controls (3.31,0.3) and (6.95,1.4) .. (10.93,3.29)   ;
\draw    (400,180) -- (400,142) ;
\draw [shift={(400,140)}, rotate = 450] [color={rgb, 255:red, 0; green, 0; blue, 0 }  ][line width=0.75]    (10.93,-3.29) .. controls (6.95,-1.4) and (3.31,-0.3) .. (0,0) .. controls (3.31,0.3) and (6.95,1.4) .. (10.93,3.29)   ;

\draw (250,123) node [anchor=north] [inner sep=0.75pt]   [align=left] {$\displaystyle a$};
\draw (370,183) node [anchor=north] [inner sep=0.75pt]   [align=left] {$\displaystyle a$};
\draw (292,93) node [anchor=north west][inner sep=0.75pt]   [align=left] {$\displaystyle b$};
\draw (308,153) node [anchor=north east] [inner sep=0.75pt]   [align=left] {$\displaystyle c$};
\draw (332,213) node [anchor=north west][inner sep=0.75pt]   [align=left] {$\displaystyle b$};
\end{tikzpicture}
}

\def\peps#1#2#3#4#5#6{
\tikzset{every picture/.style={line width=0.75pt}} 
\begin{tikzpicture}[x=0.75pt,y=0.75pt,yscale=-0.25,xscale=0.25,baseline=(current bounding box.center)]
\draw    (226,236) -- (326.23,295.62) ;
\draw    (325.77,175.62) -- (326.23,295.62) ;
\draw    (325.77,175.62) -- (226,236) ;
\draw    (226.39,336) -- (276.12,265.81) ;
\draw    (225.62,136) -- (275.89,205.81) ;
\draw    (326,235.62) -- (406,235.31) ;
\draw [color={rgb, 255:red, 208; green, 2; blue, 27 }  ,draw opacity=1 ]   (226.46,318.82) .. controls (285.2,250.8) and (256,198.71) .. (225.85,158.82) ;
\draw [color={rgb, 255:red, 208; green, 2; blue, 27 }  ,draw opacity=1 ]   (385.92,215.39) .. controls (296.02,241.93) and (275.77,175.81) .. (245.62,135.92) ;
\draw [color={rgb, 255:red, 208; green, 2; blue, 27 }  ,draw opacity=1 ]   (386.08,255.39) .. controls (312.01,237.87) and (267.2,288.04) .. (246.39,335.92) ;
\draw (223.62,133) node [anchor=south east] [inner sep=0.75pt]   [align=left] {$\displaystyle #1$};
\draw (408,235.31) node [anchor=west] [inner sep=0.75pt]   [align=left] {$\displaystyle #2$};
\draw (224.39,339) node [anchor=north east] [inner sep=0.75pt]   [align=left] {$\displaystyle #3$};
\draw (224,236) node [anchor=east] [inner sep=0.75pt]  [color={rgb, 255:red, 208; green, 2; blue, 27 }  ,opacity=1 ] [align=left] {$\displaystyle #4 $};
\draw (327.77,172.62) node [anchor=south west] [inner sep=0.75pt]  [color={rgb, 255:red, 208; green, 2; blue, 27 }  ,opacity=1 ] [align=left] {$\displaystyle #5 $};
\draw (328.23,298.62) node [anchor=north west][inner sep=0.75pt]  [color={rgb, 255:red, 208; green, 2; blue, 27 }  ,opacity=1 ] [align=left] {$\displaystyle #6 $};
\end{tikzpicture}
}

\def\MPOa#1#2#3#4#5#6{
\tikzset{every picture/.style={line width=0.75pt}} 
\begin{tikzpicture}[x=0.75pt,y=0.75pt,yscale=-0.25,xscale=0.25,baseline=(current bounding box.center)]
\draw    (380,200) -- (300,200) ;
\draw    (300,280) -- (300,200) ;
\draw    (300,280) -- (380,280) ;
\draw    (380,200) -- (380,280) ;
\draw    (420,240) -- (260,240) ;
\draw    (340,320) -- (340,160) ;
\draw [color={rgb, 255:red, 208; green, 2; blue, 27 }  ,draw opacity=1 ]   (320,160) -- (320,220) ;
\draw [color={rgb, 255:red, 208; green, 2; blue, 27 }  ,draw opacity=1 ]   (260,220) -- (320,220) ;
\draw [color={rgb, 255:red, 208; green, 2; blue, 27 }  ,draw opacity=1 ]   (260,260) -- (320,260) ;
\draw [color={rgb, 255:red, 208; green, 2; blue, 27 }  ,draw opacity=1 ]   (320,260) -- (320,320) ;
\draw [color={rgb, 255:red, 208; green, 2; blue, 27 }  ,draw opacity=1 ]   (360,260) -- (360,320) ;
\draw [color={rgb, 255:red, 208; green, 2; blue, 27 }  ,draw opacity=1 ]   (360,160) -- (360,220) ;
\draw [color={rgb, 255:red, 208; green, 2; blue, 27 }  ,draw opacity=1 ]   (360,220) -- (420,220) ;
\draw [color={rgb, 255:red, 208; green, 2; blue, 27 }  ,draw opacity=1 ]   (360,260) -- (420,260) ;
\draw (258,240) node [anchor=east] [inner sep=0.75pt]   [align=left] {$\displaystyle #1$};
\draw (340,157) node [anchor=south] [inner sep=0.75pt]   [align=left] {$\displaystyle #2$};
\draw (298,197) node [anchor=south east] [inner sep=0.75pt]  [color={rgb, 255:red, 208; green, 2; blue, 27 }  ,opacity=1 ] [align=left] {$\displaystyle #3 $};
\draw (382,197) node [anchor=south west] [inner sep=0.75pt]  [color={rgb, 255:red, 208; green, 2; blue, 27 }  ,opacity=1 ] [align=left] {$\displaystyle #4 $};
\draw (298,283) node [anchor=north east] [inner sep=0.75pt]  [color={rgb, 255:red, 208; green, 2; blue, 27 }  ,opacity=1 ] [align=left] {$\displaystyle #5 $};
\draw (382,283) node [anchor=north west][inner sep=0.75pt]  [color={rgb, 255:red, 208; green, 2; blue, 27 }  ,opacity=1 ] [align=left] {$\displaystyle #6 $};
\end{tikzpicture}
}

\def\MPOb#1#2#3#4#5#6#7{
\tikzset{every picture/.style={line width=0.75pt}} 
\begin{tikzpicture}[x=0.75pt,y=0.75pt,yscale=-0.25,xscale=0.25,baseline=(current bounding box.center)]
\draw  [fill={rgb, 255:red, 245; green, 166; blue, 35 }  ,fill opacity=1 ] (300,200) -- (380,200) -- (380,280) -- (300,280) -- cycle ;
\draw    (420,240) -- (260,240) ;
\draw    (340,320) -- (340,160) ;
\draw [color={rgb, 255:red, 208; green, 2; blue, 27 }  ,draw opacity=1 ]   (320,160) -- (320,220) ;
\draw [color={rgb, 255:red, 208; green, 2; blue, 27 }  ,draw opacity=1 ]   (260,220) -- (320,220) ;
\draw [color={rgb, 255:red, 208; green, 2; blue, 27 }  ,draw opacity=1 ]   (260,260) -- (320,260) ;
\draw [color={rgb, 255:red, 208; green, 2; blue, 27 }  ,draw opacity=1 ]   (320,260) -- (320,320) ;
\draw [color={rgb, 255:red, 208; green, 2; blue, 27 }  ,draw opacity=1 ]   (360,260) -- (360,320) ;
\draw [color={rgb, 255:red, 208; green, 2; blue, 27 }  ,draw opacity=1 ]   (360,160) -- (360,220) ;
\draw [color={rgb, 255:red, 208; green, 2; blue, 27 }  ,draw opacity=1 ]   (360,220) -- (420,220) ;
\draw [color={rgb, 255:red, 208; green, 2; blue, 27 }  ,draw opacity=1 ]   (360,260) -- (420,260) ;
\draw    (260,300) -- (340,240) ;
\draw (258,240) node [anchor=east] [inner sep=0.75pt]   [align=left] {$\displaystyle #1$};
\draw (340,157) node [anchor=south] [inner sep=0.75pt]   [align=left] {$\displaystyle #2$};
\draw (298,197) node [anchor=south east] [inner sep=0.75pt]  [color={rgb, 255:red, 208; green, 2; blue, 27 }  ,opacity=1 ] [align=left] {$\displaystyle #4 $};
\draw (382,197) node [anchor=south west] [inner sep=0.75pt]  [color={rgb, 255:red, 208; green, 2; blue, 27 }  ,opacity=1 ] [align=left] {$\displaystyle #5 $};
\draw (298,283) node [anchor=north east] [inner sep=0.75pt]  [color={rgb, 255:red, 208; green, 2; blue, 27 }  ,opacity=1 ] [align=left] {$\displaystyle #6 $};
\draw (382,283) node [anchor=north west][inner sep=0.75pt]  [color={rgb, 255:red, 208; green, 2; blue, 27 }  ,opacity=1 ] [align=left] {$\displaystyle #7 $};
\draw (258,303) node [anchor=north east] [inner sep=0.75pt]   [align=left] {$\displaystyle #3$};
\end{tikzpicture}
}

\def\vnor#1#2#3#4{
\tikzset{every picture/.style={line width=0.75pt}} 
\begin{tikzpicture}[x=0.75pt,y=0.75pt,yscale=-0.5,xscale=0.5, baseline=(current bounding box.center)]
\draw    (300,80) -- (300,62) ;
\draw [shift={(300,60)}, rotate = 450] [color={rgb, 255:red, 0; green, 0; blue, 0 }  ][line width=0.75]    (10.93,-3.29) .. controls (6.95,-1.4) and (3.31,-0.3) .. (0,0) .. controls (3.31,0.3) and (6.95,1.4) .. (10.93,3.29)   ;
\draw    (300,60) -- (300,40) ;
\draw    (260,120) -- (300,80) ;
\draw    (340,120) -- (300,80) ;
\draw    (300,200) -- (300,182) ;
\draw [shift={(300,180)}, rotate = 450] [color={rgb, 255:red, 0; green, 0; blue, 0 }  ][line width=0.75]    (10.93,-3.29) .. controls (6.95,-1.4) and (3.31,-0.3) .. (0,0) .. controls (3.31,0.3) and (6.95,1.4) .. (10.93,3.29)   ;
\draw    (300,160) -- (338.59,121.41) ;
\draw [shift={(340,120)}, rotate = 495] [color={rgb, 255:red, 0; green, 0; blue, 0 }  ][line width=0.75]    (10.93,-3.29) .. controls (6.95,-1.4) and (3.31,-0.3) .. (0,0) .. controls (3.31,0.3) and (6.95,1.4) .. (10.93,3.29)   ;
\draw    (300,180) -- (300,160) ;
\draw    (300,160) -- (261.41,121.41) ;
\draw [shift={(260,120)}, rotate = 405] [color={rgb, 255:red, 0; green, 0; blue, 0 }  ][line width=0.75]    (10.93,-3.29) .. controls (6.95,-1.4) and (3.31,-0.3) .. (0,0) .. controls (3.31,0.3) and (6.95,1.4) .. (10.93,3.29)   ;
\draw (302,60) node [anchor=west] [inner sep=0.75pt]   [align=left] {$\displaystyle #1$};
\draw (258,120) node [anchor=east] [inner sep=0.75pt]   [align=left] {$\displaystyle #2$};
\draw (302,180) node [anchor=west] [inner sep=0.75pt]   [align=left] {$\displaystyle #4$};
\draw (342,120) node [anchor=west] [inner sep=0.75pt]   [align=left] {$\displaystyle #3$};
\end{tikzpicture}
}

\def\vtcline#1{
\tikzset{every picture/.style={line width=0.75pt}} 
\begin{tikzpicture}[x=0.75pt,y=0.75pt,yscale=-0.5,xscale=0.5,baseline=(current bounding box.center)]

\draw    (320,160) -- (320,122) ;
\draw [shift={(320,120)}, rotate = 450] [color={rgb, 255:red, 0; green, 0; blue, 0 }  ][line width=0.75]    (10.93,-3.29) .. controls (6.95,-1.4) and (3.31,-0.3) .. (0,0) .. controls (3.31,0.3) and (6.95,1.4) .. (10.93,3.29)   ;
\draw    (320,120) -- (320,80) ;
\draw (322,120) node [anchor=west] [inner sep=0.75pt]   [align=left] {$\displaystyle #1$};
\end{tikzpicture}
}

\def\twovtcline#1#2{
\tikzset{every picture/.style={line width=0.75pt}} 
\begin{tikzpicture}[x=0.75pt,y=0.75pt,yscale=-0.5,xscale=0.5, baseline=(current bounding box.center)]
\draw    (306,180) -- (306,100) ;
\draw    (346,180) -- (346,100) ;
\draw (308,140) node [anchor=west] [inner sep=0.75pt]   [align=left] {$\displaystyle #1$};
\draw (348,140) node [anchor=west] [inner sep=0.75pt]   [align=left] {$\displaystyle #2$};
\end{tikzpicture}
}

\def\cpleq#1#2#3{
\tikzset{every picture/.style={line width=0.75pt}} 
\begin{tikzpicture}[x=0.75pt,y=0.75pt,yscale=-0.5,xscale=0.5, baseline=(current bounding box.center)]
\draw    (320,120) -- (300,100) ;
\draw    (320,120) -- (340,100) ;
\draw    (320,160) -- (320,120) ;
\draw    (300,180) -- (320,160) ;
\draw    (340,180) -- (320,160) ;
\draw (298,97) node [anchor=south east] [inner sep=0.75pt]   [align=left] {$\displaystyle #1$};
\draw (342,97) node [anchor=south west] [inner sep=0.75pt]   [align=left] {$\displaystyle #2$};
\draw (322,140) node [anchor=west] [inner sep=0.75pt]   [align=left] {$\displaystyle #3$};
\draw (342,183) node [anchor=north west][inner sep=0.75pt]   [align=left] {$\displaystyle #2$};
\draw (298,183) node [anchor=north east] [inner sep=0.75pt]   [align=left] {$\displaystyle #1$};
\end{tikzpicture}
}

\def\twoloop#1#2{
\tikzset{every picture/.style={line width=0.75pt}} 
\begin{tikzpicture}[x=0.75pt,y=0.75pt,yscale=-0.5,xscale=0.5,baseline=(current bounding box.center)]
\draw    (320,100) -- (311.7,91.7) -- (300,80) ;
\draw    (300,180) -- (320,160) ;
\draw    (320,160) -- (320,100) ;
\draw    (280,160) -- (300,180) ;
\draw    (260,160) -- (300,200) ;
\draw    (280,100) -- (300,80) ;
\draw    (260,160) -- (260,100) ;
\draw    (280,160) -- (280,100) ;
\draw    (300,60) -- (340,100) ;
\draw    (340,160) -- (300,200) ;
\draw    (340,160) -- (340,100) ;
\draw    (300,60) -- (260,100) ;
\draw (259,138) node [anchor=south east] [inner sep=0.75pt]   [align=left] {$\displaystyle #1$};
\draw (282,130) node [anchor=west] [inner sep=0.75pt]   [align=left] {$\displaystyle #2$};
\end{tikzpicture}
}

\def\twoloopb#1#2#3{
\tikzset{every picture/.style={line width=0.75pt}} 
\begin{tikzpicture}[x=0.75pt,y=0.75pt,yscale=-0.5,xscale=0.5,baseline=(current bounding box.center)]
\draw    (220,100) -- (240,120) ;
\draw    (240,120) -- (260,100) ;
\draw    (240,160) -- (240,120) ;
\draw    (220,180) -- (240,160) ;
\draw    (240,160) -- (260,180) ;
\draw    (220,180) -- (220,200) ;
\draw    (260,180) -- (260,200) ;
\draw    (220,200) -- (280,260) ;
\draw    (260,200) -- (280,220) ;
\draw    (280,220) -- (300,200) ;
\draw    (340,200) -- (280,260) ;
\draw    (300,200) -- (300,80) ;
\draw    (220,80) -- (220,100) ;
\draw    (280,20) -- (220,80) ;
\draw    (260,80) -- (260,100) ;
\draw    (260,80) -- (280,60) ;
\draw    (280,20) -- (340,80) ;
\draw    (280,60) -- (300,80) ;
\draw    (340,200) -- (340,80) ;
\draw (242,140) node [anchor=west] [inner sep=0.75pt]   [align=left] {$\displaystyle #3 $};
\draw (218,83) node [anchor=north east] [inner sep=0.75pt]   [align=left] {$\displaystyle #1$};
\draw (262,183) node [anchor=north west][inner sep=0.75pt]   [align=left] {$\displaystyle #2$};
\draw (218,183) node [anchor=north east] [inner sep=0.75pt]   [align=left] {$\displaystyle #1$};
\draw (262,83) node [anchor=north west][inner sep=0.75pt]   [align=left] {$\displaystyle #2$};
\end{tikzpicture}
}

\def\oneloop#1{
\tikzset{every picture/.style={line width=0.75pt}} 
\begin{tikzpicture}[x=0.75pt,y=0.75pt,yscale=-0.5,xscale=0.5,baseline=(current bounding box.center)]
\draw    (320,120) -- (300,100) ;
\draw    (280,120) -- (300,100) ;
\draw    (280,140) -- (280,120) ;
\draw    (300,160) -- (320,140) ;
\draw    (300,160) -- (280,140) ;
\draw    (320,140) -- (320,120) ;

\draw (322,123) node [anchor=north west][inner sep=0.75pt]   [align=left] {$\displaystyle #1 $};
\end{tikzpicture}
}

\begin{document}
\title{Galois conjugates of String-net Model}
\author{Chao-Yi Chen}
\affiliation{State Key Laboratory of Surface Physics, Fudan University, 200433 Shanghai, China}
\affiliation{Shanghai Qi Zhi Institute, 41st Floor, AI Tower, No. 701 Yunjin Road, Xuhui District, Shanghai, 200232, China}
\affiliation{Department of Physics and Center for Field Theory and Particle Physics, Fudan University, 200433 Shanghai, China}
\affiliation{Institute for Nanoelectronic devices and Quantum computing, Fudan University, 200433 Shanghai , China}
\author{Bing-Xin Lao}
\altaffiliation{Chen and Lao  are co- first authors of the manuscript.}
\affiliation{Department of Physics, University of Science and Technology of China}
\author{Xin-Yang Yu}
\affiliation{State Key Laboratory of Surface Physics, Fudan University, 200433 Shanghai, China}
\affiliation{Shanghai Qi Zhi Institute, 41st Floor, AI Tower, No. 701 Yunjin Road, Xuhui District, Shanghai, 200232, China}
\affiliation{Department of Physics and Center for Field Theory and Particle Physics, Fudan University, 200433 Shanghai, China}
\affiliation{Institute for Nanoelectronic devices and Quantum computing, Fudan University, 200433 Shanghai , China}
\author{Ling-Yan Hung}
\affiliation{State Key Laboratory of Surface Physics, Fudan University, 200433 Shanghai, China}
\affiliation{Shanghai Qi Zhi Institute, 41st Floor, AI Tower, No. 701 Yunjin Road, Xuhui District, Shanghai, 200232, China}
\affiliation{Department of Physics and Center for Field Theory and Particle Physics, Fudan University, 200433 Shanghai, China}
\affiliation{Institute for Nanoelectronic devices and Quantum computing, Fudan University, 200433 Shanghai , China}

\date{\today} 
\begin{abstract}
We revisit a class of non-Hermitian topological models that are Galois conjugates of their Hermitian counter parts. Particularly, these are Galois conjugates of unitary string-net models. We demonstrate that these models necessarily have real spectra, and that topological numbers are recovered as matrix elements of operators evaluated in appropriate bi-orthogonal basis, that we conveniently reformulate as a concomitant Hilbert space here. We also compute in the bi-orthogonal basis the topological entanglement entropy, demonstrating that its real part is related to the quantum dimension of the topological order. While we focus mostly on the Yang-Lee model, the results in the paper apply generally to Galois conjugates. 
\end{abstract}
\maketitle
\section{Introduction} \label{sec:Intro}
It is a fundamental premise of quantum mechanics that the Hamiltonian of a closed quantum system is Hermitian, so that the time evolution is unitary. One feature of a Hermitian Hamiltonian is that energy eigenvalues are real, which thus gives natural interpretation as energies of the eigenstates. However, Hermiticity is not a necessary condition for real eigenvalues. It is realized in the seminal paper \cite{Bender:1998gh} that non-Hermitian Hamiltonians can also acquire real eigenvalues if they are "Parity-Time-reversal"  (PT) symmetric. It is later realized that such conditions can be further relaxed and yet preserve reality of energy eigenvalues (See for example \cite{Bender:2007nj}).  Physically, such non-Hermitian Hamiltonians cannot possibly describe a closed quantum system, since they violate unitarity. They describe open systems at steady states \cite{Bender:2007nj}. These studies are mostly restricted to Gaussian systems. Interacting theories remain mostly open. 
Meanwhile, topological orders are important classes of interacting systems that go beyond the Landau-Ginzburg paradigm. Particularly in 2+1 dimensions, it have been demonstrated that these quantum phases can realize anyonic statistics that promise to realize robust universal quantum gates. Explicit (exactly solvable) lattice models that have been written down realize large classes of these models. One important class of lattice models is the Levin-Wen models. The input data is that contained in a unitary fusion category -- the unitary condition ensures that the Hamiltonian of the model is Hermitian. It is thus a curious question whether there are new topological orders if we relax the unitary condition \cite{Freedman_PRB_2012,Guo:2019fwm,Zhang_CommPhys_2020}. Clearly, this is more than pure academic curiosity, since the existence of non-Hermitian topological orders would open the possibility of designing new classes of open systems to realize fault tolerant quantum computations. 

There are already known non-unitary fusion categories that can be fed into the Levin-Wen formalism to produce exactly solvable non-Hermitian models. They are called Galois conjugates of some unitary cousins. These models have mostly been discarded as being unphysical, such as the renowned Yang-Lee model. In light of the new found interest in non-Hermiticity, the model has attracted renewed attentions. An early study of the Yang-Lee model involved embedding it in a Hermitian one is in \cite{Freedman_PRB_2012}. There are also a number of studies focused on the associated non-unitary boundary CFT \cite{2011, Lootens2020}. See also \cite{2021} for a discussion of phase transitions of the 2+1 model.  
We would like to study these examples as standalone 2+1 dimensional non-Hermitian models systematically. We found several interesting properties of this class of models. First, while it is unclear how PT symmetry is defined in these topological models, we show that Galois conjugates of unitary models have real energy spectra. This has been alluded to in \cite{Freedman_PRB_2012} but we give an explicit proof here for generic Galois duals. Second, we demonstrate that topological invariants such as the modular matrices are components of modular operators evaluated in a bi-orthogonal basis -- a notion already introduced in the non-Hermiticity literature. We use tensor network methods to construct these bi-orthogonal basis explicitly and successfully extract the modular matrices. Third, we study the entanglement entropy of the ``ground state'' wavefunctions of these non-Hermitian models using the density matrices constructed from the left/right eigenstates of the Hamiltonian. The entanglement entropy is generically complex, but we find that the real part carries a constant term that depends on the quantum dimension of the non-Hermitian topological order, analogous to the Hermitian case \cite{Levin:2006zz}.

\section{Hamiltonian construction and the reality of the spectrum}
\label{sec:Hamlitonian_construction}
\subsection{String-net model}

The basic input data of a string-net model \cite{Wen2005} is a fusion category. A category consists of a collection of objects $M$, called ``string types''. These strings could fuse, and the fusion rules are captured by the fusion coefficients $N^{c}_{ab}, $ which are non-negative integers, for $a,b,c \in M$.  The dual $j$ of a string type $i$ is one such that they fuse to the identity object. i.e. $N_{ij}^1 = 1$. 
Each string type $i$ is assigned a number called \textsl{quantum dimension} $d_i$, which satisfies $d_a d_b = \sum_c N^{c}_{ab} d_c$. We emphasise that in this paper we will relax the positive condition for $d_i$ since we are considering non-Hermitian models. Another useful variable, which we call \textsl{quantum weight}, is defined as $v_i := \sqrt{d_i}$. Since $d_i$ might be negative, $v_i$ might be pure imaginary. We also need another data called the $F$-symbol, which defines the transformation between two basis
\begin{equation}
    \ket{\Fmovebasisa{a}{b}{c}{d}{e}} = \sum_f F^{abc}_{def} \ket{\Fmovebasisb{a}{b}{c}{d}{f}} \, .
\end{equation}
In \cite{Wen2005}, $F$-symbols associated to a unitary model satisfy the pentagon equation, respects tetrahedral symmetry, and the unitary condition. In Galois conjugates of the unitary Levin-Wen models however, it is the unitary condition that is dropped. 

The Levin-Wen model is defined on a trivalent oriented graph. Each link accommodates an object from $M$, but the wavefunction of the ground state is only non-vanishing when the three objects on the three links meeting at a vertex are connected by a non-vanishing fusion coefficient. We also demand the normalization 
\begin{equation}
    \vnor{i}{k}{l}{j} = \frac{v_k v_l}{v_i} \delta_{ij}\vtcline{i} \, .
\end{equation}
In \cite{Chien-Hung_PRB_2014,Hahn_PRB_2020,Yuting_PRB_2013}, they dropped the tetrahedral symmetry condition. In \cite{Freedman_PRB_2012, Lootens2020} they dropped the unitary condition, which is similar to this paper. Another useful notation is called $G$-symbol, which is defined as $G^{a b c}_{\alpha \beta \gamma} = F^{a b \alpha}_{\beta c \gamma}/v_c v_\gamma$. 
If $F$-symbol has tetrahedral symmetry, $G$-symbol is invariant under the permutation of indices. The Hamiltonian is defined by two operators $A_v$ and $B_p$, where $v$ is the vertex and $p$ is the plaquette. $A_v$ projects the states to the subspace where the fusion rules are satisfied at each vertex. If we just focus on the ground state subspace, it can be ignored. The action of the plaquette operator $B_p = \sum_s d_s B^s_p/ \mathcal{D}$ where $\mathcal{D} := \sum_{t} d_t^2 $ and $B^s_p$ can be expressed in the following form 
\begin{equation}
\begin{aligned}\label{eq:ch4_Bps}
    &B_p^s \ket{\Bpselfdual{g}{h}{i}{j}{k}{l}} = \sum_{g_1 h_1 i_1 j_1 k_1 l_1} F^{g_1 s l}_{a g l_1} F^{h_1 s g}_{b h g_1}\\ 
    & \times F^{i_1 s h}_{c i h_1} F^{j_1 s i}_{d j i_1} F^{k_1 s j}_{e k j_1} F^{l_1 s k}_{f l k_1}
    \ket{\Bpselfdual{g_1}{h_1}{i_1}{j_1}{k_1}{l_1}} \, .
\end{aligned}
\end{equation}
To avoid clutter, we will consider models where each string type is its own dual, so the orientations on the edges cease to matter. We will also consider placing the system on a torus. We will prove that the Hamiltonian has real spectrum, though it is non-Hermitian. Since the model is topological, the reality of the spectrum would hold for lattices of arbitrary size on a torus. The simple lattice covering a torus we will consider is illustrated in the following figure. 
\begin{equation}
    \torus{a}{b}{c} := \ket{abc} \, .
\end{equation}

\subsection{Reality of the spectrum}

We inspect the spectrum of  $B_p$ operator in the subspace where fusion rules at vertices are satisfied. The form of $B_p$ is given in many places. When the lattice is the simplest one covering the torus, it is given in the supplementary material of \cite{Keren_PRL_2017}. 
\begin{equation}\label{eq:sec2_Bp}
\begin{aligned}
    &B_p \ket{abc} = \frac{1}{\mathcal{D}}\sum_{s,a',b',c'} d_s  F^{b a c'}_{s c b'} F^{c b a' }_{s a c'} F^{a c' b''}_{s b' a'} 
    \\
    &\times F^{a' c' b'}_{s b a''} F^{c' b'' a''}_{s a' c''} F^{b' a'' c''}_{s c' b''} \ket{a'' b'' c''}
    \\
    &=\frac{v_a v_b v_c v_{a''} v_{b''} v_{c''}}{\mathcal{D}} \sum_{s,a',b',c'} d_s d_{a'} d_{b'} d_{c'} 
    \times
    \\& G^{b a c}_{c' s b'} G^{c b a}_{a' s c'} G^{a c' b'}_{b'' s a'} G^{a' c' b}_{b' s a''} G^{c' b'' a'}_{a'' s c''} G^{b' a'' c'}_{c'' s b''} \ket{a'' b'' c''} \, .
\end{aligned}
\end{equation}
Using this formula we could construct the $B_p$ matrix using $a''b''c''$ as row indices and $a b c$ as column indices. In Hermitian case, because of the unitarity of the $F$-symbol, such an Hamiltonian must be Hermitian and leads to the entirely real spectrum. Since we drop the unitary condition, things are sightly different: the Hamiltonian is symmetric (it is easy to see that the matrix element is invariant under the exchange of $abc$ and $a'' b'' c''$) but not Hermitian. However, the spectrum of the Hamiltonian is still entirely real for it is a projector (the possible eigenvalue for a projector is either 1 or 0), which means that $B_p^2 = B_p$ at each plaquette $p$. In order to prove it as an projector, we need another relation, called completeness relation in \cite{Hahn_PRB_2020}:
\begin{equation}
    \twovtcline{i}{j} = \sum_{k} \frac{v_k}{ v_i v_j} N^k_{ij} \cpleq{i}{j}{k} \, .
\end{equation}
By definition, $B_p^2 = \sum_{s,t} d_s d_t B^{s}_{p} B^{t}_{p}/\mathcal{D}^2$, the operator $B^{s}_p$ is equivalent to adding a closed loop assigned with $s$ in the plaquette $p$, using graphical representation, we have 
\begin{equation}
\begin{aligned}
    B_p^2 &= \sum_{s,t} \frac{d_s d_t}{ \mathcal{D}^2}  \twoloop{s}{t} = \sum_{s,t}  \frac{d_s d_t}{ \mathcal{D}^2} \frac{v_\alpha}{v_s v_t} N^{\alpha}_{st} \twoloopb{s}{t}{\alpha} \\
    &= \sum_{s,t,\alpha} \frac{d_s d_t}{\mathcal{D}^2} \frac{v_\alpha}{v_s v_t} N^{\alpha}_{st} \frac{v_s v_t}{v_\alpha} \oneloop{\alpha}\\ 
    &= \sum_{t,\alpha} d_t  \frac{\left(\sum_{s} N^{s}_{\alpha t} d_s \right)}{\mathcal{D}^2}  \oneloop{\alpha} = \sum_{t,\alpha} \frac{d_t^2 d_\alpha}{\mathcal{D}^2} \oneloop{\alpha}  \\
    &= \sum_{\alpha}\frac{d_\alpha}{\mathcal{D}} \oneloop{\alpha} = B_p \, .
\end{aligned}
\end{equation}
Though the underlying category is still a non-unitary fusion category, the Hamiltonian continues to be a projector and its spectrum is real.

\section{Modular $S,T$ matrix for non-Hermitian Galois conjugate string-net model}
\label{sec:modular_data}

The modular $S$ and $T$ matrices are important characteristic topological data of the topological order. In Hermitian models, they can be obtained as overlaps between ground state wavefunctions before and after modular transformation of the torus.

In non-Hermitian models, the procedure needs to be re-defined, accommodating the fact that left and right eigenstates of a non-Hermitian Hamiltonian are not related by complex conjugates \cite{Weigert:2003pn,Curtright:2005zk}.  Inspired by a number of authors such as \cite{Curtright:2006aq,Curtright:2007wh,Mostafazadeh_2010}, we will introduce a \textsl{concomitant Hilbert space} that essentially accommodates a Hilbert space with a deformed inner product, which is equivalent to the bi-orthogonal basis formulation.
Though we only explicitly construct the modular matrices of the Yang-Lee model for simplicity, our method can be applied to general Galois conjugate model. To construct ground states of the Hamiltonian, we will need the central idempotents of the Yang-Lee model, which has been given in \cite{Lootens2020}.
They take the following form, 
\begin{equation}\label{eq:sec3_YL_anyon_basis}
\begin{aligned}
 &\mP_1 = (-\frac{1}{\sqrt{5} \phi'})^{\frac{3}{2}} \left(\mT^{0}_{000} + \phi' \mT_{010}^1 \right) \\
        &\mP_2 = (-\frac{1}{\sqrt{5} \phi'})^{\frac{3}{2}} \left(\mT_{111}^0 + e^{\frac{2 \pi i}{5}} \mT_{101}^{1} + \sqrt{\phi'} e^{\frac{\pi i}{5}} \mT_{111}^{1}\right) \\
        &\mP_3 = (-\frac{1}{\sqrt{5} \phi'})^{\frac{3}{2}} \left(\mT_{111}^0 + e^{\frac{-2 \pi i}{5}} \mT_{101}^{1} + \sqrt{\phi'} e^{-\frac{\pi i}{5}} \mT_{111}^{1}\right)\\
        &\mP_4 = \frac{1}{2}(-\frac{1}{\sqrt{5} \phi'})^{\frac{3}{2}} \left(\phi'^2 \mT_{000}^{0} - \phi' \mT_{010}^{1} + \phi' \mT_{111}^{0} \right.
        \\
        & \left. + \phi' \mT_{101}^{1} + \frac{1}{\sqrt{\phi'}} \mT_{111}^{1} \right),
\end{aligned}
\end{equation}
where $\mathcal{T}^{s}_{pqr}$ is the basis element of tube algebra attached with multiplication rules
\begin{equation}
\begin{aligned}
    \mT^{s}_{pqr} \mT^{s'}_{p'q'r'}     = \delta_{rp'} \sum_{f,g} F^{s'q's}_{qrg} F^{s'gp}_{s q f} F^{s g r'}_{s' q' f}\frac{v_s v_{s'}}{v_f} \mT_{pg r' }^f \, .
\end{aligned}
\end{equation}
We will show how the $T$-matrix is constructed and obtain the topological spins without direct reference to the central idempotents. Our procedure could in turn generate the central idempotent with non-trivial topological spins. 
The canonical form of the $S$-matrix has to be expressed in this anyon basis defined by the idempotent. 

\subsection{Modular $T$ matrix}
The action of modular $T$ matrix is to twist the torus.  Physically,  appropriate anyon basis -- one where an anyon loop is winding the twisted cycle, diagonalizes the $T$ matrix. 
Its eigenvalues are non-trivial phases $e^{i \theta}$ related to \textsl{topological spins} $h_i$. i.e.
\begin{equation}
    T \ket{\mP_i} = e^{i \theta_i} \ket{\mP_i}  = e^{i 2 \pi h_i}  \, ,
\end{equation}
where $\ket{\mP_i}$ is anyon basis.  Explicit wave-functions for unitary models based on a tensor network have been constructed explicitly \cite{Ran2020, Sahinoglu_JHEP_2021, Williamson_2017, Vanhove_PRL_2018, Bultinck_2017_Annual, Buerschaper_PRB_2009,Zhengcheng_PRB_2009}. 
We would like to adapt these methods to construct left/right anyon basis in the non-Hermitian model.   
Fig.~\ref{fig:sec3_tn_no_twist} illustrates the tensor network representation of a wave-function on a torus. The value of each element in the tensor network is given below:
\begin{figure}
    \centering
    \includegraphics[width=\columnwidth]{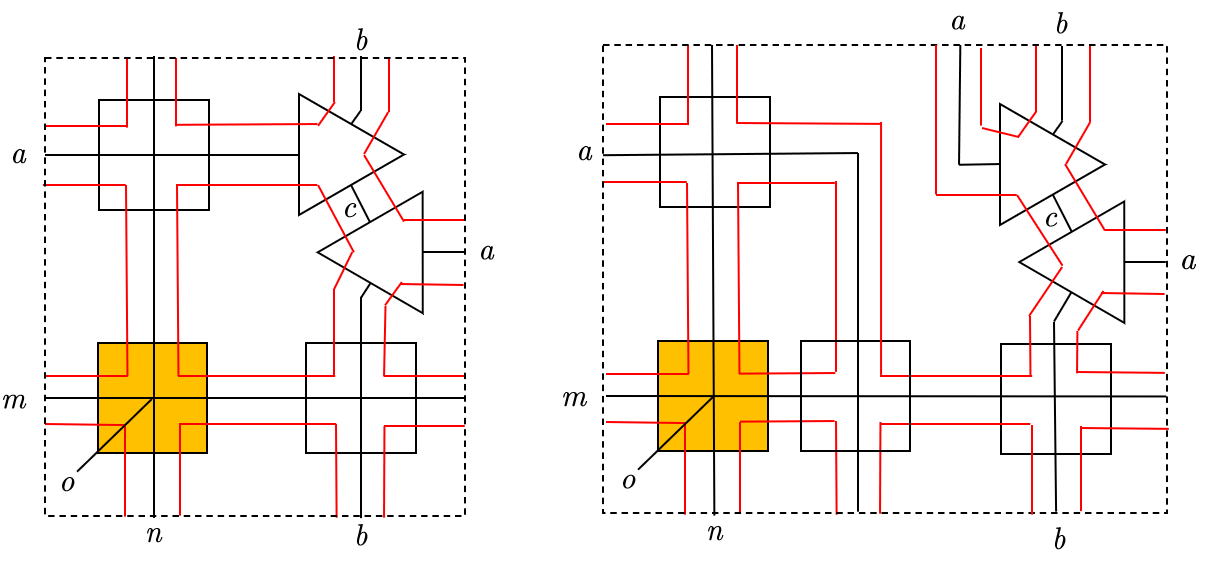}
    \caption{The tensor network representation of the string-net model ground states on a torus. The wavefunction on the right is twisted along the vertical cycle. The twist is equivalent to the modular $T$ transformation.}
    \label{fig:sec3_tn_no_twist}
\end{figure}
\begin{align}
    \peps{i}{j}{k}{\alpha}{\beta}{\gamma} &:= (v_i v_j v_k)^{\frac{1}{2}} G^{\alpha i \beta}_{j \gamma k} \, ,\\
    \MPOa{a}{i}{\alpha}{\beta}{\gamma}{\delta} &:= G^{a \gamma \alpha}_{i \beta \delta} \, ,\\
    \MPOb{m}{n}{o}{\alpha}{\beta}{\gamma}{\delta} &:= v_m v_n v_o G^{n o m}_{\gamma \alpha \beta} G^{o n m}_{\delta \beta \gamma } \, .
\end{align}
In these tensor networks,  a closed loop is further weighted by the quantum dimension. 
After contracting all the auxiliary indices of the left panel of Fig. ~\ref{fig:sec3_tn_no_twist}, we can generate the wave-function $\ket{\mT^{n}_{mom}}$. The set of such wavefunction will form an overcomplete basis $\{\ket{\mT^{n}_{mom}}\}$. We denote the wave-function after  $T$ transformation as $\ket{\mT'^{n}_{mom}}$, i.e. $T \ket{\mT^{n}_{mom}} = \ket{\mT'^{n}_{mom}}$, while the tensor network to compute this wavefunction is as the right panel of Fig.~\ref{fig:sec3_tn_no_twist}. 

\subsubsection{Detour -- bi-orthogonal basis and {\it concomitant Hilbert space}}
Computing elements of $T$ involves taking inner products of these wavefunctions with a basis bra. It is well known that eigenstates of non-Hermitian Hamiltonian of different eigenvalues are not orthogonal to each other under the definition of the usual inner product. Instead the left and right eigenstates are in fact orthogonal. They are termed bi-orthogonal basis in the literature \cite{Weigert:2003pn,Curtright:2007wh}.   
 Suppose there are two sets called right eigenstates $\{R_i\}$ and left eigenstates $\{L_j\}$, which satisfy $H \ket{R_i} = E_i \ket{R_i}$ and $H^\dagger \ket{L_i} = E_i^* \ket{L_i}$. We also require that $\bra{L_i} \ket{R_j} = \delta_{ij}$. By using this biorthogonal basis, we could expand the Hamiltonian as $H = \sum_i E_i \ket{R_i} \bra{L_i}$. Recall that the spectrum of the Galois conjugate model is all entirely real, if we define an operator called \textsl{metric operator} $g := \sum_i \ket{L_i} \bra{L_i}$ and define another Hilbert space $\mathcal{H}^c$ (we called it \textsl{concomitant Hilbert space} in order to distinguish from the original Hilbert space) equipped with a new inner product $\bra{\psi}\ket{\chi}_g := \bra{\psi} g \ket{\chi}$, we immediately find that the Hamiltonian $H$ is now Hermitian and the right eigenstates are orthonormal basis under this new inner product. 

As we are going to see, the modular matrices would reproduce their topological values when evaluated in the bi-orthogonal basis or, equivalently, in the concomitant Hilbert space with a deformed inner product. 

To construct a biorthogonal basis, we adapt the inner product $(\ket{\psi},\ket{\chi}):= \left(\ket{\psi}\right)^T \ket{\chi}$. After Gram-Schmidt orthogonalization procedure, we get two orthonormal basis $\{\ket{R_i}\}$ and $\{\ket{R'_i}\}$\footnote{Because the set $\{\ket{\mT^{n}_{mom}}\}$ is overcomplete, we can just pick the maximal linearly independent system to orthogonalize. }, which satisfy $\left(\ket{R_i},\ket{R_j}\right) = \delta_{ij}$, $\left(\ket{R'_i},\ket{R'_j}\right) = \delta_{ij}$ and $T \ket{R_i} = \ket{R'_i}$. It is easy to see that the corresponding left eigenstate of $\ket{R_i}$(or $\ket{R'_i}$) is just the conjugation of itself. The inner product $\bra{\cdot}\ket{\cdot}_g$ is now $(\cdot, \cdot)$. 

Now we are ready to evaluate the components of the $T$ matrix in the concomitant Hilbert space.  Choosing the set $\{\ket{R_i}\}$ as the representation, the matrix element is given by
\begin{equation}\label{eq:sec3_T_matrix}
    T_{ij} := \bra{R_i} \ket{T R_j}_g = \bra{R_i}\ket{R'_j}_g = \left(\ket{R_i},\ket{R'_j}\right) \, .
\end{equation}
By using this scheme we could construct the $T$ matrix under the basis $\{\ket{R_i}\}$. Though it is not diagonal, we could still read the topological spins by finding its eigenvalue. For example, in Yang-Lee model, the topological spins corresponding to the anyon basis~(\ref{eq:sec3_YL_anyon_basis}) are
\begin{equation}
    h_1 =h_4= 0,\quad h_2 = -\frac{2}{5},\quad h_3 = \frac{2}{5} \, .
\end{equation}
For the non-trivial anyon basis (with non-trivial topological spins), we can directly get them because its eigensubspace is one dimensional. In the appendix~\ref{appendix:T_matrix_property}, we prove that such a representation is a real representation of $T$ matrix. This property immediately leads that the conjugation of the anyon basis with non-trivial twist is also the anyon basis. Besides, another important conclusion is that the anyon basis will automatically form the biorthogonal basis, which will be used in the calculation of entanglement entropy.

\subsection{Modular $S$ matrix}
The $S$ matrix carries information of braiding between anyons. Under the action of the $S$ transform, the lattice covering the torus is rotated by $90^{\circ}$. 
We have already constructed the wavefunction of the basis of the tube algebra by using Fig.~\ref{fig:sec3_tn_no_twist}, which we denote by $\{\ket{\mT^{n}_{mom}}\}$. We label the wave-function after the $S$ transformation as $\{\ket{\mS^{n}_{mom}}\}$. The algorithm to compute it is to rotate the MPO (the orange box in Fig.~\ref{fig:sec3_tn_no_twist}) and use F-moves to transform it back into its original form, i.e.
\begin{equation}
    S \ket{\mT^{n}_{mom}} = \ket{\mS^{n}_{mom}} = \sum_{o'} F^{nmn}_{mo o'} \ket{\mT^{m}_{no'n}} \, .
\end{equation}
Again, the set $\{\ket{\mT^{n}_{mom}}\}$ and $\ket{\mS^{n}_{mom}}$ are two overcomplete sets of basis. After choosing their maximal linearly independent system, we orthogonalize these two sets of basis as in the computation of $T$ matrix, denoting them as $\{\ket{R_i}\}$ and $\{\ket{R^{S}_i}\}$. The construction of $S$ matrix is also similar:
\begin{equation}
    (S_0)_{ij} := \bra{R_i} \ket{S R_j}_g = \bra{R_i}\ket{R^S_j}_g = \left(\ket{R_i},\ket{R^S_j}\right) \, .
\end{equation}
However, if we want to get the $S$ matrix which satisfies the Verlinde formula, we need to transform the representation. The final $S$ matrix is $S = U^\dagger S_0 U$, where $U$ is the unitary transformation relating the basis $\ket{R_i}$ and the anyon basis. For example, using our construction and the anyon basis in Eq.~(\ref{eq:sec3_YL_anyon_basis}), the $S$ matrix for Yang-Lee model is 
\begin{equation}
    S = \frac{1}{\mathcal{D}} \left(
    \begin{matrix}
     1 & -1/\phi \\
     -1/\phi & -1 
    \end{matrix}
    \right) \otimes \left(
    \begin{matrix}
     1 & -1/\phi \\
     -1/\phi & -1 
    \end{matrix}
    \right) \, .
\end{equation}

\section{Entanglement Entropy of non-Hermitian system}\label{sec:entropy}

Entanglement entropy is one important quantity characterizing a topological order. In a non-Hermitian system, density matrices have been constructed using the left/right eigenstates of the Hamiltonian, and the entanglement entropy is computed from these generalized density matrices  \cite{Poyao2020,Kawabata:2020olo,Chen:2020bhm,Xi:2020iji, Ohlsson_PRA_2021,Tu:2021xje}. These prior studies focused mostly on free systems. We would like to apply these methods to compute the entanglement entropy and recover also the topological term characterizing the non-Hermitian topological order. 

\begin{figure}
    \centering
    \includegraphics[width=0.7\columnwidth]{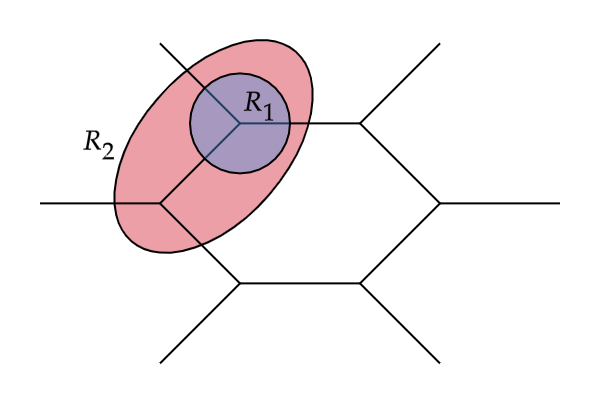}
    \caption{The sub-regions we choose on a torus in the computation of the entanglement entropies.}
    \label{fig:sec4_entropy}
\end{figure}

Explicitly, the density operator is defined as $\rho_G := \rho g$ \cite{Poyao2020,Kawabata:2020olo,Chen:2020bhm,Xi:2020iji, Ohlsson_PRA_2021,Tu:2021xje}, where $g$ is the metric operator we introduced in the previous section. Let us continue to take Yang-Lee model as an example. Consider the ground state on a torus with the anyon with topological spin $h_2 = - 2/5$ wrapping a non-contractible cycle and we denote the state as $\ket{\psi_2}$. The corresponding generalized density operator is $\rho_G = \ket{\psi_2}\bra{\psi_2} g$. We assume that the state can be expanded by using $\{\ket{R_i}\}$, i.e. $\ket{\psi_2} = \sum_i \beta_2^i \ket{R_i}$. We can simplify the equation 
\begin{equation}
\begin{aligned}
    \rho_G &= \sum_{i,j} \beta_2^i (\beta_2^j)^* \ket{R_i} \bra{R_j} \left(\sum_k \ket{L_k} \bra{L_k}\right) \\
    &= \sum_{i,j} \beta_2^i (\beta_2^j)^* \ket{R_i} \bra{L_j} \, ,
\end{aligned}
\end{equation}
where we have used the bi-orthogonal relation. As we discussed in the last section and the  appendix~\ref{appendix:T_matrix_property}, the left eigenstate $\ket{L_k} = \ket{R_k}^*$. 
To simplify the expression further, 
consider the central idempotent corresponding to $h_3 = 2/5$. Its coefficient $\beta_3^j$ satisfies $\beta_3^j = (\beta_2^j)^*$. The above density operator finally reduces to $\rho = \ket{\psi_2} (\ket{\psi_3})^T $, where  $\ket{\psi_3}$ is the state associated to a winding anyon with spin $h_3=2/5$ generated by the corresponding idempotent. 
Consider the reduced entanglement entropy of region $R_1$ and $R_2$ as labeled in Fig.~(\ref{fig:sec4_entropy}).
One can verify that the entropy for arbitrary regions (for example, the region $R_1$) strictly obeys the formula given in \cite{Levin_PRL_2006},
\begin{equation}
    S_R = -\sum_{\{X\}} \frac{\prod_m d_{q_m}}{\mathcal{D}^{n-1}} \log \left(\frac{\prod_l d_{q_l}}{\mathcal{D}^{n-1}}\right) \, ,
\end{equation}
where $\{X\}$ denotes all the possible string structure in region $R$, $q_m$ denotes the leg crossing the boundary of $R$ and $n$ denotes the number of the legs crossing the boundary. 
Noticing that some of the quantum dimensions are negative, the log function can not be simplified in the same way as in \cite{Levin_PRL_2006}. 
The entanglement entropy is generically complex. The real part of the topological entanglement entropy however is similar to the usual Hermitian case,
\begin{equation}
    \Re \left[S_{\text{top}}\right] = - \log \mathcal{D} \, ,
\end{equation}
recovering the quantum dimension of the non-Hermitian model.

\section{Conclusion}
In this paper, we revisit the Galois conjugates of unitary string net models. These models are epitome of non-Hermitian systems that demonstrate topological properties. 
While it is demonstrated that they cease to be topological when embedded into Hermitian systems, in this paper we study these models as they are -- and demonstrated that topological quantities are embedded in operators computed in bi-orthogonal basis, or equivalently, the ``concomitant'' Hilbert space with a deformed inner product. 
Specifically, quantities like the modular matrices and entanglement entropies recover topological invariants, exactly as their Hermitian counterpart. 
This is in contrast with previous works \cite{Freedman_PRB_2012} which observed that keeping only the right (left) eigenstates and embed them in a Hermitian model destroys the topological properties of the model. 

For interacting models like topological orders, it is not entirely clear how time reversal symmetry should be defined. Therefore it is not immediately clear how one should explain the reality of the spectrum, even though we prove this generically in Galois conjugated Levin-Wen models. 

Much more is needed to understand non-Hermitian topological orders and how to utilize their topological robustness in open systems in light of the concomitant Hilbert space. For example, we would expect that expectation values computed via the deformed inner product should correspond to the expectation values of operators actually measured in a steady state, which can be checked in experiments.
These are important questions that we will hopefully revisit in the near future.


\section*{Acknowledgements} \label{sec:acknowledgements}
LYH acknowledges the support of NSFC (Grant
No. 11922502, 11875111) and the Shanghai Municipal Science and Technology Major Project
(Shanghai Grant No.2019SHZDZX01). We thank Lin Chen, Ce Shen, Jiaqi Lou, Xiangdong Zeng, Xirong Liu, for insightful discussions related to this work.

\appendix*
\section{Data of general $su(2)_k$ model} \label{appendix:G_symbol_product}
For completeness and preparing for the backgrounds for the proof below, we review the computation of the $F$-symbol and $G$-symbol of $su(2)_k$ model along the lines adopted in \cite{aasen2020topological}. The associated category is called $\mathcal{A}_{k+1}$ and its simple objects are labelled by $0,\frac{1}{2},1,\frac{3}{2}, \ldots , \frac{k}{2}$. The fusion rule for these objects are $N^a_{bc} = 1$ if $a+b \ge c, \, b+c\ge a, \, c+a\ge b, \, a+b+c \in \mathbb{Z},$ and $a+b+c \ge k$, and otherwise $N^{a}_{bc}$ = 0. It is easy to verify that all the integer element (or we called integer spin) will form a sub-algebra, which we used as examples in this paper. The $G$ symbol is computed by the following formula,
\begin{equation}
    G^{abc}_{def} = \left\{
    \begin{matrix}
    a &b &c\\
    d &e &f
    \end{matrix}
    \right\}_q (-1)^p \, , 
\end{equation}
where 
\begin{widetext}
\begin{equation}
p := \frac{3(a+b+c+d+e+f)^2 - (a+d)^2 - (b+e)^2 - (c+f)^2}{2}
\end{equation}
Since we only focus on the integer object, when $a,b,c,d,e,f$ are all integers, $p$ must be an integer. The other part is called $q$-deformed Wigner-6j symbols, it is computed via the Racah formula,
\begin{equation}
    \begin{aligned}
    \left\{
    \begin{matrix}
    j_1 &j_2 &j_3\\
    j_4 &j_5 &j_6
    \end{matrix}
    \right\}_q &= \Delta(j_1,j_2,j_3) \Delta(j_1,j_5,j_6) \Delta(j_4, j_2,j_6) \Delta(j_4, j_5,j_3) \\
    &\times \sum_{z}\frac{(-1)^z [z+1]!}{[z-a_1]! [z- a_2]! [z-a_3]! [z-a_4]! [b_1-z]! [b_2 - z]! [b_3 - z]!} \, .
    \end{aligned}
\end{equation}
where
\begin{equation}
     q := e^{\frac{2g \pi i}{k+2}}, \quad
     [m] := \frac{q^{m/2} - q^{-m/2}}{q^{1/2}-q^{-1/2}} = \frac{\sin \left(\frac{g m \pi}{k + 2}\right)}{\sin \left(\frac{g \pi }{k+2}\right)}, \quad [n]! := \left\{
     \begin{aligned}
     \prod_{m=1}^n [m] &\quad n > 0 \\
     1                &\quad n=0 \\
     \infty           &\quad n\le 0
     \end{aligned}
     \right. \, .
\end{equation}
The summation $\sum_{z}$ is over integers $z$ such that every term in the summand is well-defined, i.e. the summation is from $\max \{a_1,a_2,a_3,a_4\}$ to $\min\{b_1, b_2, b_3\}$ where $a_1 = j_1 + j_2 + j_3$, $a_2 = j_1 + j_5 + j_6$, $a_3 = j_4 + j_2 + j_6$, $a_4 = j_4 + j_5 +j_3$, $b_1 = j_1+j_2+j_4+j_5$, $b_2 = j_2+j_3+j_5+j_6$ and $b_3 = j_3+j_1+j_6+j_4$. The rest term is the function $\Delta(j_1,j_2,j_3)$, it is defined as 
\begin{equation}
    \Delta(j_1, j_2, j_3) := \left\{
    \begin{aligned}
    &\frac{([j_1 + j_2 - j_3]!)^{1/2} ([j_3 + j_1 - j_2]!)^{1/2} ([j_2 + j_3 - j_1]!)^{1/2}}{([j_1 + j_2 + j_3 + 1]!)^{1/2}} \quad & N^{j_1}_{j_2 j_3} = 1\\
    &0 \quad &\text{otherwise}
    \end{aligned}
    \right. \, .
\end{equation}
The quantum dimension of the simple object $x$ is $d_x = [2 x + 1]$. The integer factor $g$ is a new parameter compared to the formula in \cite{aasen2020topological}. Galois conjugation is to change the factor $g$, expanding the allowed value  $g=1$ by $1,
\ldots, \{k/2\}$, where $\{a\}$ denotes the integer part of number $a$. For example, Fibonacci model is in fact the subset of integer spin of $su(2)_3$ model, where we set $g = 1$ and deformation parameter $q = e^{\frac{2 \pi i}{5}}$. If we set $g = 2$, $q = e^{\frac{4 \pi i}{5}}$, we obtain the Yang-Lee model. The non-unitarity comes from the possible negative sign of the function $[\cdots]$. The $F$-symbol can be computed by the formula $F^{a b \alpha}_{\beta c \gamma} = v_c v_\gamma G^{a b c}_{\alpha \beta \gamma}$. 

Let us further consider these $su(2)_k$ models with integer spins. One important observation is that the summation must be a real number. The only possible imaginary part comes from the function like $\Delta(j_1, j_2 ,j_3)$. In the following discussion, we will make use of the observation that for arbitrary combination of $a,b,c,m,n,o,\alpha_1,\alpha_2,\alpha_3,\alpha_4$ the product of the following $G$-symbol must be real, i.e.
\begin{equation}
     G^{\alpha_1 b \alpha_2}_{c \alpha_4 a} G^{\alpha_4 c \alpha_2}_{a \alpha_3 b}  G^{a \alpha_3 \alpha_2}_{n \alpha_1 \alpha_4} G^{m \alpha_1 \alpha_4}_{b \alpha_3 \alpha_2} G^{n o m}_{\alpha_2 \alpha_3 \alpha_4} G^{o n  m}_{\alpha_1 \alpha_4 \alpha_2} \in \mathbb{R} \quad\text{for all possible cases} \, .
\end{equation}
The proof of this observation more generally is direct. Since the only possible imaginary term is $\Delta(*,*,*)$, we denote each term in the product as 
\begin{equation}
\begin{aligned}
    G^{\alpha_1 b \alpha_2}_{c \alpha_4 a} &\propto \Delta(c,b,a)\Delta(c,\alpha_4,\alpha_2)\Delta(\alpha_1,b,\alpha_2)\Delta(\alpha_1,\alpha_4,a) \, ,\\
    G^{\alpha_4 c \alpha_2}_{a \alpha_3 b} &\propto \Delta(a,c,b)\Delta(a,\alpha_3,\alpha_2) \Delta(\alpha_4,c,\alpha_2)\Delta(\alpha_4,\alpha_3,b) \, ,\\
    G^{a \alpha_3 \alpha_2}_{n \alpha_1 \alpha_4} &\propto
    \Delta(a,\alpha_1,\alpha_4) \Delta(b,\alpha_3,\alpha_4)\Delta(m,\alpha_1,\alpha_4)\Delta(m,\alpha_3,\alpha_2) \, ,\\
    G^{m \alpha_1 \alpha_4}_{b \alpha_3 \alpha_2} &\propto
    \Delta(b,\alpha_1,\alpha_2)\Delta(b,\alpha_3,\alpha_4)\Delta(m,\alpha_1,\alpha_4)\Delta(m,\alpha_3,\alpha_2) \, , \\
    G^{n o m}_{\alpha_2 \alpha_3 \alpha_4}  &\propto
    \Delta(\alpha_2,\alpha_3,m)\Delta(n,o,m)\Delta(n,\alpha_3,\alpha_4)\Delta(\alpha_2,o,\alpha_4)\, , \\
    G^{o n  m}_{\alpha_1 \alpha_4 \alpha_2} &\propto
    \Delta(\alpha_1,\alpha_4,m)\Delta(o,n,m)\Delta(\alpha_1,n,\alpha_2)\Delta(o,\alpha_4,\alpha_2) \, .
\end{aligned}
\end{equation}
It is easy to see that $\Delta(i,j,k) = \Delta(j,i,k) = \Delta(k,i,j)$, each $\Delta$ appears in pairs and $\Delta^2$ must be real. We then finish the proof. 
\end{widetext}

\section{$T$ matrix property} \label{appendix:T_matrix_property}
In this appendix we will prove several important properties of the $T$ matrix and anyon basis. First, the $T$ matrix remains unitary, i.e. $T^{-1} = T^{\dagger}$. The second theorem we want to prove is that under the basis $\{\ket{R_i}\}$, $T$ matrix is in addition a real matrix. Before we prove this theorem, we need several lemmas. Recall that the wave-function $\{\ket{\mT^{n}_{mom}}\}$ and $\{\ket{\mT'^{n}_{mom}}\}$ are generated by the left/right panel of Fig.~\ref{fig:sec3_tn_no_twist} separately. If we write down the wave-function explicitly,
\begin{widetext}
\begin{align}
    \ket{\mT^{n}_{mom}} &= \sum_{abc} v_a v_b v_c v_m v_n v_o\sum_{\alpha_1,\alpha_2,\alpha_3,\alpha_4}  \left(\prod_{i}^4 d_{\alpha_i}\right)
    G^{\alpha_1 b \alpha_2}_{c \alpha_4 a} G^{\alpha_4 c \alpha_2}_{a \alpha_3 b}  G^{a \alpha_3 \alpha_2}_{n \alpha_1 \alpha_4} G^{m \alpha_1 \alpha_4}_{b \alpha_3 \alpha_2} G^{n o m}_{\alpha_2 \alpha_3 \alpha_4} G^{o n  m}_{\alpha_1 \alpha_4 \alpha_2} \ket{abc} \, , \\
    \ket{\mT'^{n}_{mom }} &= \sum_{abc} v_a v_b v_c v_m v_n v_o \sum_{\alpha_i} \left(\prod_{j}d_{\alpha_j}\right) G^{\alpha_1 b \alpha_2}_{c \alpha_4 a} G^{\alpha_4 c \alpha_2}_{a \alpha_3  b} G^{a \alpha_3 \alpha_2}_{n \alpha_4 \alpha_5} G^{m \alpha_4 \alpha_5}_{a \alpha_4 \alpha_1} G^{m \alpha_1 \alpha_4}_{b \alpha_3 \alpha_2} G^{n o m}_{\alpha_2 \alpha_3 \alpha_5} G^{o n m}_{\alpha_4 \alpha_5 \alpha_2} \ket{abc} \, .
\end{align}
\end{widetext}
Similar to the expression~(\ref{eq:sec2_Bp}), the product of $G$ symbol is real, which can be verified by following the procedure in the last appendix~\ref{appendix:G_symbol_product}. Immediately we obtain the fact below: the inner product of two basis (for arbitrary combination)  takes the following form
\begin{align}
    \left(\ket{\mT^{n}_{mom}}, \ket{\mT^{n_1}_{m_1 o_1 m_1}}\right) = v_{m}v_{m_1} v_{n} v_{n_1} v_{o} v_{o_1} B^{m n o m_1 n_1 o_1}\, , \\
    \left(\ket{\mT'^{n}_{mom}}, \ket{\mT'^{n_1}_{m_1 o_1 m_1}}\right) = v_{m}v_{m_1} v_{n} v_{n_1} v_{o} v_{o_1} B'^{m n o m_1 n_1 o_1}\, , \\
    \left(\ket{\mT^{n}_{mom}}, \ket{\mT'^{n_1}_{m_1 o_1 m_1}}\right) = v_{m} v_{m_1} v_{n} v_{n_1} v_{o} v_{o_1} C^{mno m_1 n_1 o_1}\, , \label{eq:appendix3_innerproduct}
\end{align}
where the factors $B^{m n o m_1 n_1 o_1}$, $B'^{m n o m_1 n_1 o_1}$ and $C^{mno m_1 n_1 o_1}$ are all real. The set $\{\ket{\mT^{n}_{mom}}\}$ is complete\footnote{Here we just pick the maximal linearly independent system.}, there must exist a unique combination of the coefficient $\alpha^j_{mno}$ and $\alpha'^{j}_{mno}$ such that
\begin{align}
    \ket{R_j}  = \sum_{m,n,o} \alpha^{j}_{mno} \ket{\mT^{n}_{mom}} \, , \\
    \ket{R'_j} = \sum_{m,n,o} \alpha'^{j}_{m n o }\ket{\mT'^{n}_{mom}} \, .
\end{align}
The second lemma we want to prove is that $\alpha^{j}_{mno} \propto v_m v_n v_o$ and $\alpha'^{j}_{mno} \propto v_m v_n v_o$. Because the procedures to prove these two cases are totally the same, we just need to prove that $\alpha^{j}_{mno}$ satisfies the condition. In order to simplify our notations, we relabel the indices $m,n,o$, and define $\mathbf{v}_i := \ket{\mT^{n_i}_{m_i o_i m_i}}$. The Gram-Schmidt orthogonalization process can be explicitly written as the following form 
\begin{widetext}
\begin{equation}\label{eq:app3_orthogonalization}
    \ket{R_j}  = \frac{1}{D_{j-1}} \left|
    \begin{matrix}
        (\mathbf{v}_1,\mathbf{v}_1) &(\mathbf{v}_2,\mathbf{v}_1) & \ldots &(\mathbf{v}_j,\mathbf{v}_1) \\
        (\mathbf{v}_1,\mathbf{v}_2) &(\mathbf{v}_2,\mathbf{v}_2) & \ldots &(\mathbf{v}_j,\mathbf{v}_2) \\
        \vdots &\vdots &\ddots &\vdots \\
        (\mathbf{v}_1,\mathbf{v}_{j-1}) &(\mathbf{v}_2,\mathbf{v}_{j-1}) & \ldots &(\mathbf{v}_j,\mathbf{v}_{j-1}) \\
        \mathbf{v}_1 &\mathbf{v}_2 &\ldots &\mathbf{v}_j
    \end{matrix}
    \right| \, ,
\end{equation}
where the factor $D_{j-1}$ is 
\begin{equation}
    D_{j} = \left|
    \begin{matrix}
        (\mathbf{v}_1,\mathbf{v}_1) &(\mathbf{v}_2,\mathbf{v}_1) & \ldots &(\mathbf{v}_j,\mathbf{v}_1) \\
        (\mathbf{v}_1,\mathbf{v}_2) &(\mathbf{v}_2,\mathbf{v}_2) & \ldots &(\mathbf{v}_j,\mathbf{v}_2) \\
        \vdots &\vdots &\ddots &\vdots \\
        (\mathbf{v}_1,\mathbf{v}_{j}) &(\mathbf{v}_2,\mathbf{v}_{j}) & \ldots &(\mathbf{v}_j,\mathbf{v}_{j})
    \end{matrix}
    \right| \,  
\end{equation}
with $D_0 = 1$.
\end{widetext}
First we need to prove that $D_j$ is real for all $j$. Using the Levi-Civita symbol, the determinant can be expanded as
\begin{equation}
    D_j = \sum_{i_1,\ldots i_j= 1}^j \epsilon_{i_1 i_2 \ldots i_j } (\mathbf{v}_1,\mathbf{v}_{i_1}) \cdots  (\mathbf{v}_j,\mathbf{v}_{i_j}) \, .
\end{equation}
for each label $i_j$, there must exist a positive integer that equals to it, which means that there must exist two copies of $v_{m_{i_j}}v_{n_{i_j}}v_{o_{i_j}}$. All the possible imaginary part will be canceled and thus the resultant $D_j$ must be real. The same trick applies to Eq.~(\ref{eq:app3_orthogonalization}). Expanding the result 
\begin{equation}
\begin{aligned}
    \ket{R_j} &= \frac{1}{D_{j-1}} \sum_{i_1,\ldots,i_j = 1}^j \epsilon_{i_1,\ldots,i_j}  \\
    & \times (\mathbf{v}_1,\mathbf{v}_{i_1}) (\mathbf{v}_2,\mathbf{v}_{i_2}) \cdots (\mathbf{v}_{j-1},\mathbf{v}_{i_{j-1}} ) \mathbf{v}_{i_j}
\end{aligned}
\end{equation}
Again, in the term $(\mathbf{v}_1,\mathbf{v}_{i_1}) (\mathbf{v}_2,\mathbf{v}_{i_2}) \cdots (\mathbf{v}_{j-1},\mathbf{v}_{i_{j-1}} ) \mathbf{v}_{i_j}$, for $i_k \neq i_j$, there must exist a positive integer $k$ that equals to $i_k$, which will cancel the possible imaginary part (the product of the form $v_m v_n v_o$). The imaginary part only comes from $\left(\mathbf{v}_{i_j},\mathbf{v}_l\right)$. It will leave the possible imaginary part $v_{m_{i_j}} v_{n_{i_j}} v_{o_{i_j}}$, which completes the proof that $\alpha^{j}_{mno} \propto v_m v_n v_o$. Recall that the construction of the $T$ matrix Eq.~(\ref{eq:sec3_T_matrix})
\begin{equation}
\begin{aligned}
    T_{ij} &= \left(\ket{R_i},\ket{R'_j}\right) \\
    &=\sum_{m,n,o,m_1,n_1,o_1} \alpha^{i}_{mno} \alpha'^{j}_{m_1 n_1 o_1} \left(\ket{\mT^{n}_{mom}}, \ket{\mT'^{n_1}_{m_1 o_1 m_1}}\right).
\end{aligned}
\end{equation}
Using the conclusion $\alpha^{j}_{mno} \propto v_m v_n v_o$ and the Eq.~(\ref{eq:appendix3_innerproduct}), we can see that all the possible imaginary parts will cancel exactly and thus $T_{ij}$ must be a real number as claimed.  

Several interesting facts follow. We denote the eigenstate as $\ket{\mP^R_{\lambda_i}}$, where the capital $R$ denotes the right eigenstate and $\lambda_i$ is its corresponding eigenvalue, i.e. $T \ket{\mP^R_{\lambda_i}} = \lambda_i \ket{\mP^R_{\lambda_i}}$. Since $T$ matrix is a real matrix, we automatically have 
\begin{equation}
    T^* \ket{\mP^{R}_{\lambda_i}}^* = \lambda_i^* \ket{\mP^{R}_{\lambda_i}}^* = T \ket{\mP^{R}_{\lambda_i}}^* = \lambda_i^* \ket{\mP^R_{\lambda_i^*}}\, .
\end{equation}
The last equality means that $\ket{\mP^{R}_{\lambda_i}}^*$ is also an eigenstate, which corresponds to the eigenvalue $\lambda_i^*$. Particularly, noticing that $\lambda_i$ must be the root of unity (Vafa's theorem), if $\lambda_i \neq 1$, we could generate the dual anyon basis by conjugation. Another important property is that the corresponding left eigenstate of the anyon basis $\ket{\mP^{R}_{\lambda_i}}$ is itself, i.e. $\ket{\mP^{L}_{\lambda_i}} = \ket{\mP^{R}_{\lambda_i}}$. Recalling the definition of the biorthogonal system, what we need to prove is that $\bra{\mP^{R}_{\lambda_i}}\ket{\mP^{R}_{\lambda_j}} = \delta_{\lambda_i \lambda_j}$. The proof relies on the unitarity of the $T$ matrix. 
\begin{equation}\label{eq:app3_biortho_proof}
\begin{aligned}
    \bra{\mP^{R}_{\lambda_i}}\ket{\mP^{R}_{\lambda_j}} &= \ket{\mP^{R}_{\lambda_i}}^\dagger \ket{\mP^{R}_{\lambda_j}} \\
    &= \frac{1}{\lambda_j} \ket{\mP^{R}_{\lambda_i}}^\dagger \lambda_j \ket{\mP^{R}_{\lambda_j}} = \frac{1}{\lambda_j} \ket{\mP^{R}_{\lambda_i}}^\dagger T \ket{\mP^{R}_{\lambda_j}} \\
    &= \frac{1}{\lambda_j} \left(T^\dagger  \ket{\mP^{R}_{\lambda_i}}\right)^\dagger 
    \ket{\mP^{R}_{\lambda_j}}
\end{aligned}
\end{equation}
Since $T$ is unitary, $T^\dagger$ is its inverse. 
Also, $\lambda_i^{-1} = \lambda_i^*$ because $\lambda_i$ must be the root of unity. We have
\begin{equation}
    T \ket{\mP^{R}_{\lambda_i}} = \lambda_i \ket{\mP^R_{\lambda_i}} \, \Rightarrow \, T^\dagger \ket{\mP^{R}_{\lambda_i}} = \lambda_i^* \ket{\mP^R_{\lambda_i}} \, .
\end{equation}
Then Eq.~(\ref{eq:app3_biortho_proof}) can be simplified into 
\begin{equation}
    \frac{1}{\lambda_j} \left(T^\dagger  \ket{\mP^{R}_{\lambda_i}}\right)^\dagger 
    \ket{\mP^{R}_{\lambda_j}} = \frac{\lambda_i}{\lambda_j} \bra{\mP^{R}_{\lambda_i}}\ket{\mP^{R}_{\lambda_j}}
\end{equation}
If $\lambda_i \neq \lambda_j$, we directly have $\bra{\mP^{R}_{\lambda_i}}\ket{\mP^{R}_{\lambda_j}} = 0$. If $\lambda_i  = \lambda_j$, we can normalize the anyon basis to let $\bra{\mP^{R}_{\lambda_i}}\ket{\mP^{R}_{\lambda_i}} = 1$. One might worry about degeneracies, but we argue that we can always orthogonalize the eigenstates with the same eigenvalue in the corresponding eigensubspace.

\bibliography{reference} 
\end{document}